\documentclass[twocolumn,fleqn,12pt]{article} 
\usepackage[margin= 1 in]{geometry}
\usepackage{amsmath}
\usepackage{mathtools}
\usepackage{physics}
\usepackage{graphicx}
\usepackage{xcolor}
\usepackage{caption}
\usepackage{subcaption}
\usepackage{cuted}
\usepackage[utf8]{inputenc}
\usepackage{pdfpages}
\usepackage{cite}
\usepackage{natbib}
\usepackage{authblk}
\usepackage{float}

\begin{document}
\title{Study of Duffing oscillator using an improved Lindstedt-Poincaré method and relevant comparisons}

\author[1]{Ramij Ahamed}
\author[1]{Subhankar Ray\thanks{Email: subhankar.ray@jadavpuruniversity.in}}
\affil[1]{Department of Physics, Jadavpur University, Kolkata, India}

\maketitle

\begin{abstract}
\noindent
The undamped Duffing oscillator is a nonlinear dynamical system with broad applications in physics, engineering and biological system. We present a comprehensive analysis of this system using the Lindstedt-Poincaré method (LPM) and its modifications and make comparison with numerical solution obtained using higher order Runge-Kutta. It is also shown the method suggested in this article converges better than the standard LPM and Lindstedt-Poincaré method with Burton's modification.
\end{abstract}

\noindent
\rule{.95\linewidth}{0.1pt} \\


\section{Introduction:}
Nonlinear oscillatory behaviour occurs widely in both natural and engineered systems, ranging from mechanical vibrations to oscillations in electrical circuits and various rhythmic changes biological systems \cite{strogatz2001nonlinear}. The Duffing oscillator is one of the most important models in nonlinear dynamics. It first appeared in a small book by Georg Duffing in 1918, where he summarized his systematic investigation of various mechanical nonlinear oscillators \cite{duffing1918erzwungene}. The Duffing oscillator displays rich dynamical features and has been extensively studied\cite{nayfeh2024nonlinear}.
We consider here the normalized undamped, unforced Duffing equation given by,
\begin{equation}
\frac{\mathrm{d^2}x(t)}{\mathrm{d}t^2}+ x(t)+\epsilon x(t)^3=0
\label{eq:1}  
\end{equation} 
with the initial conditions, $x(0)=x_0$ and $\dot{x}(0) = v_0$. For $\epsilon>0$, the system behaves as a hardening spring, whereas for $\epsilon<0$ it exhibits softening spring behaviour. Several perturbation methods like Regular Perturbation, Lindstedt-Poincaré method (LPM) \cite{Lindstedt1883}, Multiple Time scale\cite{}, homotopy perturbation\cite{feng2009homotopy}, He's energy balance\cite{momeni2011application} and He's parameter-expansion\cite{xu2007he} etc have been applied to the analyze Duffing oscillator\cite{nayfeh2024perturbation, jordan2007nonlinear}. In this article, we introduce a Modified Lindstedt-Poincaré Method (LPM-M) and compare it with the standard Lindstedt-Poincaré Method (LPM) and Burton's improvement of LPM (LPM-B)\cite{burton1984perturbation}. 
In standard literature, the solution of Duffing oscillator using LPM and LPM-B is usually presented upto third order.
In this work, we propose a modified LPM to solve the Duffing oscillator. The LPM-M shows remarkably good convergence by fourth order which is matched by LPM and
LPM-B only after calculations upto 10th and 8th orders respectively.

\section{Analysis by Lindstedt-Poincaré Method:}
The LPM is one of the most frequently used analytical techniques for studying nonlinear oscillator systems that involve a small perturbation parameter \cite{nayfeh2024perturbation}.

In LPM, the frequency ($\omega $) is taken to be dependent on the perturbation parameter ($\epsilon$), rather than a constant. By introducing a stretched time variable,  $\tau=\omega  t$, \eqref{eq:1} can be rewritten as
\begin{equation}
\omega^2  x''(\tau)+x(\tau)+\epsilon  x(\tau)^3=0, 
\label{eq:2}   
\end{equation}
where ($'$) denotes differentiation with respect to $\tau$. In standard LPM,  $\omega$ and $x(\tau)$ are expanded in a power series in $\epsilon$ upto $N$th order

\begin{eqnarray}
\omega_{[N]} = \sum_{i=0}^{N} \epsilon^i \omega_i , \;\   x_{[N]}(\tau) = \sum_{i=0}^{N} \epsilon^i x_i(\tau) \label{eq:4}
\end{eqnarray}

Inserting $\omega_{[N]}$ and $x_{[N]}(\tau)$ in \eqref{eq:2}, we get the following set of equations
\begin{flalign}
O(\epsilon^0):x_0''+x_0&=0 \label{eq:5}&&\\
O(\epsilon^1):x_1''+x_1&=- x_0^3-2 \omega_1  x_0'' \label{eq:6}&&\\
O(\epsilon^2):x_2''+x_2 &=- 3x_0^2 x_1-\omega_1^2  x_0''-2 \omega_2  x_0'' &&\nonumber\\
&-2 \omega_1 x_1'' \label{eq:7} &&\\
O(\epsilon^3):x_3''+x_3&=-3 x_0 x_1^2-3 x_0^2 x_2-2 \omega_1 \omega_2  x_0''&&\nonumber\\
&-2 \omega_3  x_0''-\omega_1^2 x_1''-2  \omega_2 x_1''&&\nonumber\\
&-2 \omega_1  x_2'' \label{eq:8} &&
\end{flalign}

\noindent
\begin{flalign}
O(\epsilon^4): x_4''+x_4&=- x_1^3-6x_0 x_1 x_2- 3x_0^2 x_3&&\nonumber\\
&-\omega_2^2 x_0''-2 \omega_1 \omega_3  x_0''-2 \omega_4  x_0''&&\nonumber\\
&-2  \omega_1 \omega_2  x_1''-2 \omega_3  x_1''-\omega_1^2  x_2''&&\nonumber\\
&-2 \omega_2 x_2''-2 \omega_1 x_3' \label{eq:9} && \\
O(\epsilon^5):x_5''+x_5&=-3x_1^2 x_2-3x_0 x_2^2-6x_0 x_1 x_3 && \nonumber\\
&-3x_0^2 x_4-2 \omega_2 \omega_3 x_0-2 \omega_1 \omega_4 x_0''&&\nonumber\\
&-2 \omega_5 \ddot x_0-\omega_2^2 x_1''-2 \omega_1 \omega_3 x_1''&&\nonumber\\
&-2 \omega_4 x_1''-2 \omega_1 \omega_2 x_2''-2 \omega_3 \ddot x_2&&\nonumber\\
&-\omega_1^2 x_3''-2 \omega_2 x_3''-2 \omega_1 x_4'' \label{eq:10}&&
\end{flalign}
...
\begin{flalign}
O(\epsilon^{10}): x_{10}''+x_{10}&=- x_3^3+...-6x_2 x_3 x_4  \label{eq:11}     &&
\end{flalign}

The goal is to solve the equation with initial condition $x(0)=x_0$ and $x'(0)=v_0$. Here we shall consider $x(0)=A$ and $v_0=0$. 
The initial conditions become
\begin{align} 	
&x_0(0)=A,\;\; x_i(0)=0 \; ( i=1,..,10)\\
&x_0'(0)=0, \hspace*{0.35cm} x_i'(0)=0 \; ( i=1,..,10).
\end{align}

\noindent
From \eqref{eq:5}, using the relevant initial conditions, we get
\begin{equation}
x_0(\tau)=A \cos\tau
\label{eq:13}
\end{equation}

\noindent
After putting this $x_0(\tau)$ in \eqref{eq:6} we get
\begin{equation}
x_1''+x_1=\left(2A\omega_1-\frac{3A^3}{4}\right)\cos\tau-\frac{A^2}{4} \cos3\tau \label{eq:14}
\end{equation}

In order to remove resonance that leads to unbounded solutions, the coefficient of $\sin(\tau)$ and $\cos(\tau)$ are independently set to zero. This process is called removal of secular terms. Thus the first order frequency correction is
\begin{equation}
\omega_1=\frac{3A^2}{8} \label{eq:15}
\end{equation}

\noindent
and \eqref{eq:14} simplifies to
\begin{equation}
x_1''+x_1=-\frac{A^2}{4} \cos3\tau \label{eq:16}
\end{equation}
\noindent
which yields the 1st order solution
\begin{equation}
x_1(\tau)=\frac{A^3}{32} (-\cos\tau+\cos3\tau ) \label{eq:17}
\end{equation}

\noindent
After replacing $x_0$, $\omega_1$ and $x_1$ in \eqref{eq:7} we obtain,
\begin{equation}
\begin{split}
x_2''+x_2=\left(2A\omega_2+\frac{21A^5}{128}\right)\cos\tau\\
+\frac{3A^5}{16} \cos3\tau-\frac{3A^5}{128} \cos5\tau \label{eq:18}
\end{split}
\end{equation}

Following similar procedure of removing secular terms, we get the 2nd order frequency correction and 2nd order solution

\noindent
\begin{flalign}
 \omega_2 =&-\frac{21A^4}{256}, \nonumber &&\\
x_2(\tau) &= \frac{A^5}{1024 }(23 \cos\tau-24 \cos3\tau +\cos5\tau ) &&
\label{eq:19}
\end{flalign}

\noindent
Considering successive orders, we obtain,

\begin{flalign}
\omega_3 =&\frac{81A^6}{2048}, \nonumber && \\
x_3(\tau)&=\frac{A^7}{32768 }(-547 \cos\tau+594 \cos3\tau \nonumber &&\\
& -48\cos5\tau+\cos7\tau) \label{eq:20} && 
\end{flalign}

%

\noindent
\begin{flalign}
\omega_4 =&-\frac{6549A^8}{262144} , \nonumber && \\
x_4(\tau)&=\frac{A^9}{1048576 }(13426 \cos\tau-15121 \cos3\tau \nonumber &&\\
&+1766 \cos5\tau-72 \cos7\tau+\cos9\tau) \label{eq:21} && \\
\omega_5 =& \frac{37737A^{10}}{2097152} , \nonumber &&\\
x_5(\tau) &= \frac{A^{11}}{33554432 }(-339176 \cos\tau+394701 \nonumber &&\\
&  \cos3\tau-58944 \cos5\tau +3514 \cos7\tau \nonumber &&\\
& -96 \cos9\tau+\cos11\tau) \label{eq:22} &&\\
&...&&\\
\omega_{10} =& -\frac{497158650207 a^{20}}{70368744177664} , &&  \nonumber\\
x_{10}(\tau)&= \frac{A^{21}}{1125899906842624 } \cdot && \nonumber \\
(&-4821475754931 \cos\tau+...+\cos21\tau) \label{eq:22.1} &&
\end{flalign}


\noindent
Hence, one obtains

\begin{eqnarray}
\omega_{[10]} &=& \sum_{i=0}^{10} \epsilon^i \omega_i \nonumber \\
x_{[10]}(\tau) &=& \sum_{i=0}^{10} \epsilon^i x_i(\tau) \nonumber 
\end{eqnarray}

where we replace $\omega_i$ and $x_i(\tau)$ from expressions obtained above.
\begin{strip}
\noindent\rule{\textwidth}{0.4pt}
\begin{flalign}
\omega_{[10]} &= 1+\frac{3A^2}{8} \epsilon -\frac{21A^4}{256}\epsilon^2+ \frac{81A^6}{2048}\epsilon^3-\frac{6549A^8}{262144}\epsilon^4 +\frac{37737A^{10}}{2097152}\epsilon^5-\frac{936183 A^{12}}{67108864}\epsilon^6  +\frac{6077907 A^{14}}{536870912} \epsilon^7 \nonumber && \\
& \hspace*{2cm} -\frac{2604833685 A^{16}}{274877906944}  \epsilon^8+\frac{17839453041 A^{18}}{2199023255552} \epsilon^9-\frac{497158650207 A^{20}}{70368744177664} \epsilon^{10} \nonumber  &&
\end{flalign}
\end{strip}


%
%
%
\section{Analysis by LPM-B:}
A more accurate result for frequency $\omega$ can be found by expanding $\omega^2$, rather than $\omega$, in a power series in $\epsilon$. This expansion is superior  because we are dealing with a second order differential equation \cite{burton1984perturbation}.
\\

In order to differentiate calculation using this method we shall use a stretched time variable  $\tau_1=\alpha t$, \eqref{eq:1} can be rewritten as
\begin{equation}
\alpha^2  z''(\tau_1)+z(\tau_1)+\epsilon  z(\tau_1)^3=0, \label{eq:23}  
\end{equation}
where ($'$) denotes differentiation with respect to $\tau_1$. In this method, the $\alpha^2$ and $z(\tau_1)$ are expanded in a power series in $\epsilon$ upto $N$th order
\begin{equation}
\alpha^2_{[N]} = \sum_{i=0}^{N} \epsilon^i \alpha_i, \;\  z_{[N]}(\tau_1) = \sum_{i=0}^{N} \epsilon^i z_i(\tau_1) \label{eq:25}
\end{equation}

\noindent
After removing secular terms in different orders of $\epsilon$, get frequencies and solutions,
\begin{flalign}
z_0(\tau_1)&=A \cos \text{$\tau_1$} && \label{eq:26} \\
\alpha_1=&\frac{3A^2}{4}, \nonumber && \\
z_1(\tau_1)&=\frac{A^3}{32} (-\cos\tau_1+\cos3\tau_1 ) && \label{eq:27} 
\end{flalign}


\noindent
\begin{flalign}
\alpha_2=&-\frac{3A^4}{128}, \nonumber &&\\
z_2(\tau_1)&=\frac{A^5}{1024}(23 \cos\tau_1-24 \cos3\tau_1+\cos5\tau_1 ) && \label{eq:28}
\end{flalign}
\begin{flalign}
\alpha_3= & \frac{9A^6}{512}, \nonumber && \\
z_3(\tau_1)&=\frac{A^7}{32768}(-547 \cos\tau_1+594 \cos3\tau_1 \nonumber && \\
	&-48\cos5\tau_1+\cos7\tau_1) && \label{eq:29}
\end{flalign}
\begin{flalign}
\alpha_4= & -\frac{1779A^8}{131072}, \nonumber && \\
z_4(\tau_1)&=\frac{A^9}{1048576}(13426 \cos\tau_1-15121 \cos3\tau_1 \nonumber &&\\
	&+1766 \cos5\tau_1-72 \cos7\tau_1+\cos9\tau_1) && \label{eq:30} 
\end{flalign}
\begin{flalign}
\alpha_5= & \frac{5643A^{10} }{524288} \nonumber  && \\
z_5(\tau_1)&= \frac{A^{11}}{33554432}(-339176 \cos\tau_1+394701 \nonumber && \\
	& \cos3\tau_1-58944 \cos5\tau_1 +3514 \cos7\tau_1 \nonumber && \\
	&-96 \cos9\tau_1+\cos11\tau_1) && \label{eq:31} \\
	& \dots \dots \dots \nonumber && \\
\alpha_8=& -\frac{841910643 A^{16}}{137438953472} \nonumber  &&  \\
z_8(\tau_1)& = \frac{A^{17}}{1099511627776 }(6297662471 \cos\tau_1+ ... \nonumber && \\ 
	& ...+\cos17\tau_1) 
\end{flalign}
\noindent
Substituting we get
\begin{equation*}
\begin{split}
\alpha^2_{[8]}=& 1+\frac{3A^2}{4} \epsilon-\frac{3A^4}{128}\epsilon^2+ \frac{9A^6}{512}\epsilon^3-\frac{1779A^8}{131072}\epsilon^4 \\
& +\frac{5643A^{10}}{524288}\epsilon^5-\frac{146661 A^{12}}{16777216} \epsilon^6 \\
& +\frac{486603 A^{14}}{67108864} \epsilon^7  -\frac{841910643 A^{16}}{137438953472} \epsilon^8
\end{split} 
\end{equation*}

\noindent
So, we get $\alpha_{[8]}$

\begin{strip}
\noindent\rule{\textwidth}{0.4pt}
\begin{equation*}
\alpha_{[8]} =
\sqrt{
1+\frac{3A^2}{4} \epsilon-\frac{3A^4}{128}\epsilon^2+ \frac{9A^6}{512}\epsilon^3-\frac{1779A^8}{131072}\epsilon^4 +...+\frac{486603 A^{14}}{67108864} \epsilon^7-\frac{841910643 A^{16}}{137438953472} \epsilon^8
}       
\end{equation*}
\end{strip}

\noindent
From \eqref{eq:25}
\begin{eqnarray}
z_{[8]}(\tau_1) &=& \sum_{i=0}^{8} \epsilon^i z_i(\tau_1) \nonumber
\end{eqnarray}

where $z_i(\tau_1)$ are inserted from expressions obtained above.

\section{Analysis by Modified LPM:}
In Modified LPM, the exact  frequency ($\omega_{ex}$) is calculated using the property of so called turning points.
This method is expected to give better result than the LPM and LPM(Burton) as in these cases truncated frequencies are used in each step of perturbation. The frequency and
amplitude both keeps improving as higher and higher order calculations are made. For the Duffing oscillator it is possible to evaluate exact frequency ($\omega_{ex}$) valid for all orders.

%
%
%
\subsection{Exact Time Period:}
Using the first integral of \eqref{eq:1} , we get the energy equation \citep{cveticanin2018strong}
\begin{equation}
\left(\frac{\mathrm{d}x(t)}{\mathrm{d}t}\right)^2+ x(t)^2+\frac{1}{2} \epsilon x(t)^4 =C
\end{equation}

\noindent
Using the initial conditions, we have $C=A^2+ \frac{1}{2} \epsilon A^4$. As the potential is symmetric with $x$, we can get the time period by rearranging

\begin{flalign}
\;\ \;\ T = 4 \int_{0}^{A} \frac{dx}{\sqrt{\left( A^2-x^2 \right) \left[ 1+\frac{1}{2}\epsilon(A^2+x^2)\right]}} && \label{eq:37}
\end{flalign}

\noindent
After putting $x=A \sin \theta$,  in \eqref{eq:37}

\begin{align}
T &= 4 \sqrt{\frac{2}{2+p}} \cdot \int_{0}^{\frac{\pi}{2}}  \frac{d\theta}{\sqrt{1+m_1 \sin^2 \theta}} && \nonumber \\
&= 4 \sqrt{\frac{2}{2+p}} \cdot F \left(\frac{\pi}{2}|-m_1 \right) && \label{eq:37.1}
\end{align}

\noindent
where, $p=\epsilon A^2$, $m_1=\frac{p}{2+p} $ and $F(\phi | m )$ is the incomplete elliptic integral of the first kind defined as
\begin{equation}
F(\phi | m )= \int_{0}^{\phi} \frac{d\theta}{\sqrt{1-m \sin^2 \theta}}  \nonumber
\end{equation}

\noindent
Using the formula \citep{abramowitz1964handbook}
\begin{equation}
\begin{split}
F(\phi | -m )= \frac{1}{\sqrt{1+m}} K \left( \frac{m}{1+m} \right) \\
- \frac{1}{\sqrt{1+m}} F\left(\frac{\pi}{2}-\phi | \frac{m}{1+m} \right)
\end{split} \nonumber
\end{equation}

\noindent
we get
\begin{equation}
F\left(\frac{\pi}{2} | -m_1 \right)= \frac{1}{\sqrt{1+m_1}} K \left( \frac{m_1}{1+m_1} \right)  \nonumber
\end{equation}

\noindent
where, $K(m)$ is the complete elliptic integral of the first kind defined as
\begin{equation}
K( m )= \int_{0}^{\frac{\pi}{2}} \frac{d\theta}{\sqrt{1-m \sin^2 \theta}}  \nonumber
\end{equation}

\noindent
Using the above in \eqref{eq:37.1}

\begin{align}
T &= 4 \sqrt{\frac{2}{2+p}} \cdot  \frac{1}{\sqrt{1+m_1}} K \left( \frac{m_1}{1+m_1} \right) \nonumber \\
&= \frac{4}{\sqrt{1+p}}  K \left( \frac{p}{2(1+p)} \right) \label{eq:38}
\end{align}

\subsection{Finding $x$ order by order in modified LPM:}
We see that in the L-P method, determining the solution upto $kth$ order allows the frequency to be corrected upto the same $k^{th}$ ($\omega_k$) term. In this approach, the exact frequency is inserted in the perturbation solution and it is expanded and rearranged to remove secular terms. Now, we will apply the modified L-P method in the undamped Duffing oscillator and solve upto $5^{th}$ order solution. From \eqref{eq:38}, the angular frequency
\begin{eqnarray}
\omega_{ex}=\frac{2\pi}{T}
=\frac{\pi}{2} \cdot  \frac{ \sqrt{1+p}}{K \left( \frac{p}{2(1+p)}\right)} \label{eq:39}
\end{eqnarray}

\noindent
Now we expand $f(p)=\frac{ \sqrt{1+p}}{K \left( \frac{p}{2(1+p)}\right)} $ in power series of $p$ about $p=0$.
\begin{equation}
f(p)=\sum_{n=0}^{\infty} a_n p^n
\end{equation}
Using the standard procedure, 
\begin{equation}
a_n= \frac{1}{n!} \lim_{p \to 0} \frac{d^n f(p)}{dp^n}
\ 
\end{equation}
we get $a_n$
\begin{flalign}
&  a_0=\frac{2}{\pi },a_1=\frac{3}{4 \pi },a_2=-\frac{21}{128 \pi }, a_3=\frac{81}{1024 \pi }, \nonumber &&\\
& \;\ \;\ a_4=-\frac{6549}{131072 \pi }, \;\ a_5=\frac{37737}{1048576 \pi }, ...  \nonumber &&
\end{flalign}
So, we have
\begin{flalign}
f(p) =& \frac{2}{\pi }+\frac{3 p}{4 \pi }-\frac{21 p^2}{128 \pi }+\frac{81 p^3}{1024 \pi }-\frac{6549 p^4}{131072 \pi } \nonumber   && \\
& +\frac{37737 p^5}{1048576 \pi }+ ... \nonumber &&
\end{flalign}
\noindent
and
\begin{flalign}
\omega_{ex} = &  1+\frac{3 p}{8}-\frac{21 p^2}{256}+\frac{81 p^3}{2048}-\frac{6549 p^4}{262144}\nonumber && \\
 & \;\ \;\ \;\  +  \frac{37737 p^5}{2097152} + ... && \nonumber\\
= & 1+\frac{3A^2 \epsilon}{8} -\frac{21A^4 \epsilon^2}{256}+ \frac{81A^6 \epsilon^3}{2048} \nonumber && \\
& \;\ \;\  -\frac{6549A^8 \epsilon^4}{262144}  +\frac{37737A^{10} \epsilon^5}{2097152}+ ... && \nonumber\\
=& \sum_{i=0}^{N} \epsilon^i \nu_i+O(\epsilon^{N+1}) \nonumber \\
= &  \nu_{[N]}+O(\epsilon^{N+1}) \label{eq:42}
\end{flalign}

\noindent
We see that $\nu_0=1,\nu_1=\omega_1, \nu_2=\omega_2, \nu_3=\omega_3, \nu_4=\omega_4, \nu_5=\omega_5$ and so on. By introducing a stretched time variable, $\tau_2=\omega_{ex} t$,\eqref{eq:1} can be rewritten as
\begin{equation}
\nu_{[N]}^2 \ddot{y}(\tau_2)+ y(\tau_2)+\epsilon y(\tau_2)^3=0 ,  \label{eq:43}
\end{equation}
where ( $\dot{•}$ ) denotes differentiation with respect to $\tau_2$. Here, we put the the truncated frequency ($\nu_{[N]}$) in place of exact frequency ($\omega_{ex}$) to remove the secular terms. Now $y(\tau_2)$ is expanded in a power series in $\epsilon$ upto $N$th order
\begin{eqnarray}
y_{[N]}(\tau_2) &=& \sum_{i=0}^{N} \epsilon^i y_i(\tau_2) \label{eq:44} 
\end{eqnarray}

\noindent
Inserting  $y_{[N]}(\tau_2)$ in \eqref{eq:43}, we get the following set of equations

\noindent
\begin{flalign}
O(\epsilon^0):\ddot{y_0}+y_0&=0&& \label{eq:45} \\
O(\epsilon^1):\ddot{y_1}+y_1 &=-y_0^3-2 \nu_1  \ddot{y_0} && \label{eq:46}\\
O(\epsilon^2):\ddot{y_2}+y_2 &= -3y_0^2 y_1-\nu_1^2  \ddot{y_0}-2 \nu_2  \ddot{y_0} && \nonumber\\
&-2 \nu_1 \ddot{y_1} && \label{eq:47} \\
O(\epsilon^3):\ddot{y_3}+y_3 &=-3 y_0 y_1^2-3 y_0^2 y_2-2 \nu_1 \nu_2  \ddot{y_0} && \nonumber\\
&-2 \nu_3  \ddot{y_0}-\nu_1^2 \ddot{y_1}-2 \nu_2 \ddot{y_1} && \nonumber\\
&-2 \nu_1  \ddot{y_2} && 
\end{flalign}
\begin{flalign}
O(\epsilon^4): \ddot{y_4}+y_4 &=- y_1^3-6y_0 y_1 y_2- 3y_0^2 y_3 && \nonumber\\
&-\nu_2^2 \ddot{y_0}-2 \nu_1 \nu_3  \ddot{y_0}-2 \nu_4  \ddot{y_0} && \nonumber\\
&-2  \nu_1 \nu_2  \ddot{y_1} -2 \nu_3  \ddot{y_1}-\nu_1^2  \ddot{y_2} && \nonumber\\
&-2 \nu_2 \ddot{y_2}-2 \nu_1 \ddot{y_3} && \\
O(\epsilon^5):\ddot{y_5}+y_5 &=-3y_1^2 y_2-3y_0 y_2^2+6y_0 y_1 y_3 && \nonumber\\
&-3y_0^2 y_4-2 \nu_2 \nu_3 y_0-2 \nu_1 \nu_4 \ddot{y_0} && \nonumber\\
&-2 \nu_5 \ddot{y_0}-\nu_2^2 \ddot{y_1}-2 \nu_1 \nu_3 \ddot{y_1} && \nonumber\\
&-2 \nu_4 \ddot{y_1}-2 \nu_1 \nu_2 \ddot{y_2}-2 \nu_3 \ddot{y_2} && \nonumber\\
&-\nu_1^2 \ddot{y_3}-2 \nu_2 \ddot{y_3} -2 \omega_0 \nu_1 \ddot{y_4} &&
\end{flalign}

\noindent
and the initial conditions
\begin{eqnarray}
y_0(0)&=&A, \;\  y_i(0)=0 \;\ ( i=1,,5)\\
\dot{y_0}(0)&=&0, \;\  \dot{y_i}(0)=0 \;\ ( i=1,,5)
\end{eqnarray}

\noindent
From \eqref{eq:45}, using the relevant initial conditions, we get
 \begin{equation}
y_0(\tau_2)=A \cos\tau_2
\end{equation}

\noindent
After putting this $y_0(\tau_2)$ in \eqref{eq:46} we get
\begin{equation}
\ddot{y_1}+y_1=-\frac{A^2}{4} \cos3\tau_2 \label{eq:54}
\end{equation} 
\noindent
There is no resonance terms in \eqref{eq:54}. Using the relevant initial conditions we get 
\begin{eqnarray}
y_1(\tau_2)=\frac{A^3}{32} (-\cos\tau_2+\cos3\tau_2 )
\end{eqnarray}

\noindent
After putting this $y_0(\tau_2)$ and $y_1(\tau_2)$ in \eqref{eq:47} we get
\begin{equation}
\ddot{y_2}+y_2=\frac{3A^5}{128} (8\cos3\tau_2-\cos5\tau_2)
\end{equation}

\noindent
Using the relevant initial conditions we get
\begin{flalign}
& y_2(\tau_2)=\frac{A^5}{1024}(23 \cos\tau_2-24 \cos3\tau_2+\cos5\tau_2 ) &&
\end{flalign}

\noindent
Considering successive orders, we obtain
\begin{flalign}
 y_3(\tau_2)=& \frac{A^7}{32768}(-547 \cos\tau_2+594 \cos3\tau_2 \nonumber &&\\
&-48\cos5\tau_2+\cos7\tau_2) &&\\
y_4(\tau_2)=&\frac{A^9}{1048576}(13426 \cos\tau_2-15121 \cos3\tau_2 \nonumber &&\\
&+1766 \cos5\tau_2-72 \cos7\tau_2+\cos9\tau_2) &&\\
y_5(\tau_2)=& \frac{A^{11}}{33554432}(-339176 \cos\tau_2+394701 \nonumber &&\\
& \cos3\tau_2 -58944 \cos5\tau_2 +3514 \cos7\tau_2  \nonumber && \\
&-96 \cos9\tau_2+\cos11\tau_2) &&
\end{flalign}

\noindent
From \eqref{eq:44}
\begin{eqnarray}
y_{[5]}(\tau_2) &=& \sum_{i=0}^{5} \epsilon^i y_i(\omega_{ex} t) \label{eq:44} 
\end{eqnarray}

\section{Results and Discussion:}
\begin{figure}[h]
\centering
\includegraphics[width=.9\linewidth]{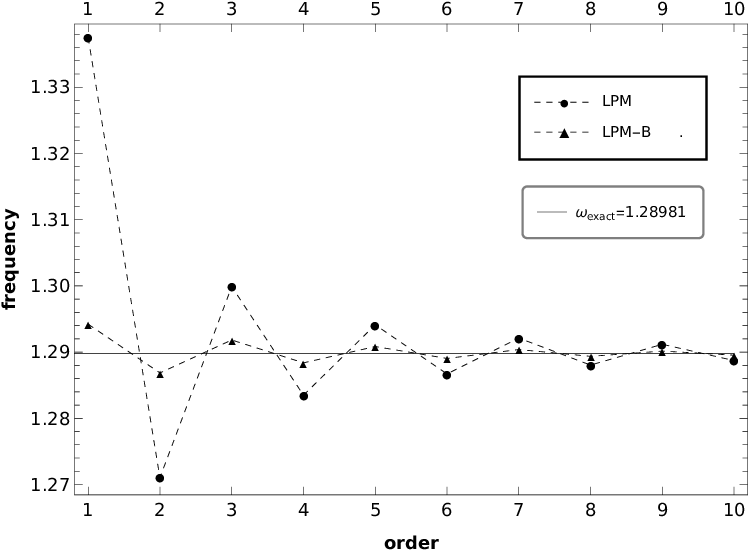}
\captionof{figure}{Frequency for different orders of LPM and LPM-B}
\label{fig:1}
\end{figure}

\noindent
The standard LPM, LPM-B and the LPM-M suggested in this paper, are compared with high precision numerical data to understand the efficacy and relative merits of these methods.
The comparison was performed for various values of perturbation parameter $\epsilon$ and amplitude $A$. In the figures presented here, the perturbation parameter and amplitude are taken to be $\epsilon =0.4$ and $A=1.5$. The exact frequency and the time period are obtained as $ \omega = 1.28981$ and $T= 4.8714$.
In our LPM-M, for a given $\epsilon$, the exact frequency $\omega_{ex}$ is used in all orders of calculation. However, in standard LPM and LPM-B, the frequencies $\omega$ approach the exact frequency in an oscillatory manner, as the order of calculation is increased. When compared with the standard LPM, in LPM-B, the approach to $\omega_{ex}$ is faster as shown in Figure \ref{fig:1}.

\begin{figure}[h]
	\centering
	\includegraphics[width=.9\linewidth]{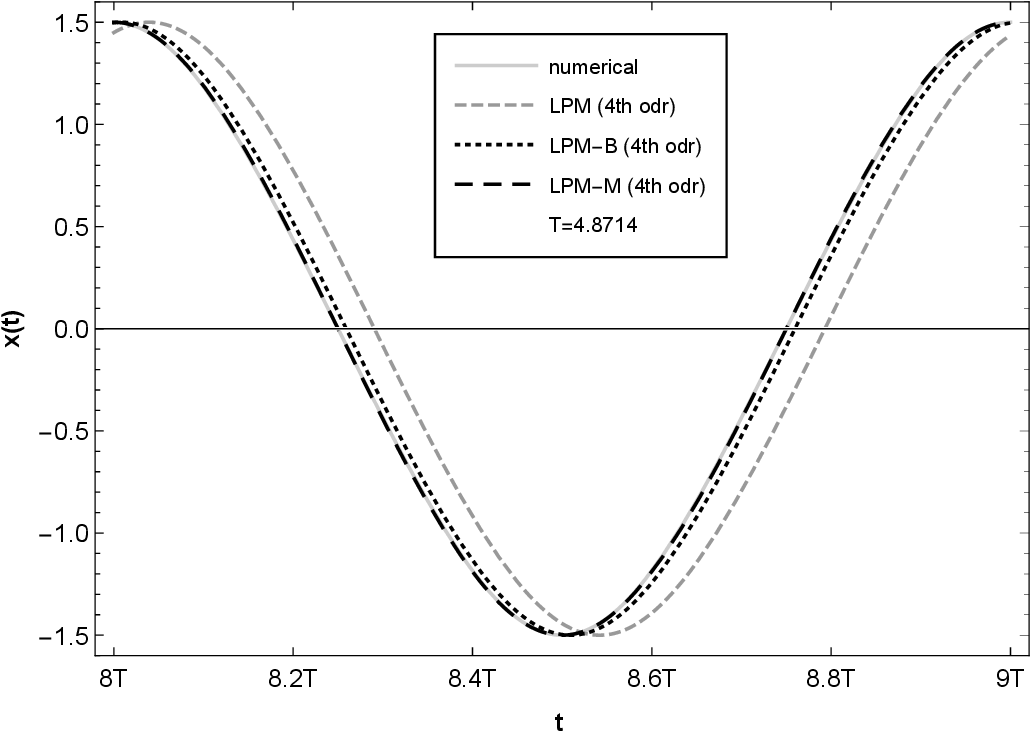}
	\captionof{figure}{$x_{[4]}(t)$ comparison in lower time $(8T\leq t \leq 9T)$}
	\label{fig:2}
\end{figure}

\begin{figure}[h]
	\centering
        \includegraphics[width=.9\linewidth]{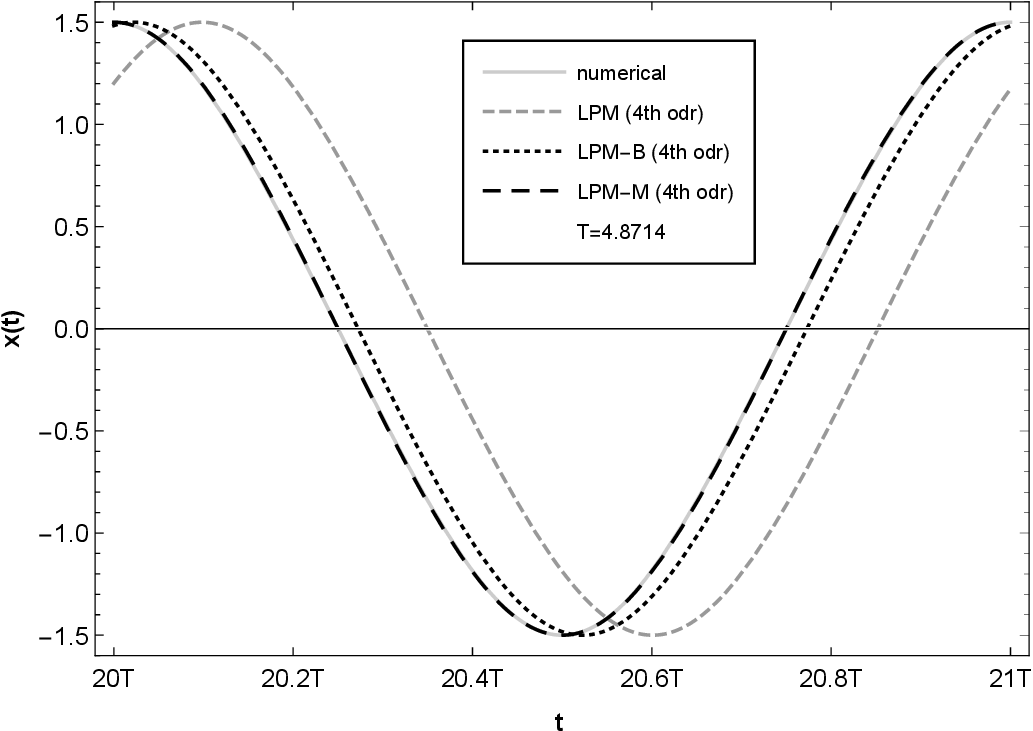}
\captionof{figure}{$x_{[4]}(t)$ comparison in higher time $(20T\leq t \leq 21T)$}
\label{fig:3}
\end{figure}

Figure \ref{fig:2} and Figure \ref{fig:3} show the Fourth-order solutions obtained with the three methods together with the high precision numerical solution for small $t$, $8 T\le t \le 9 T$ and large $t$,$20 T\le t \le 21 T$ . It is observed that standard LPM and Burton's modified LPM solutions show a phase shift, which increases for larger $t$, whereas the LPM-M solutions remain in phase with the numerical solution. 

\begin{figure}[h]
        \centering
\includegraphics[width=\linewidth]{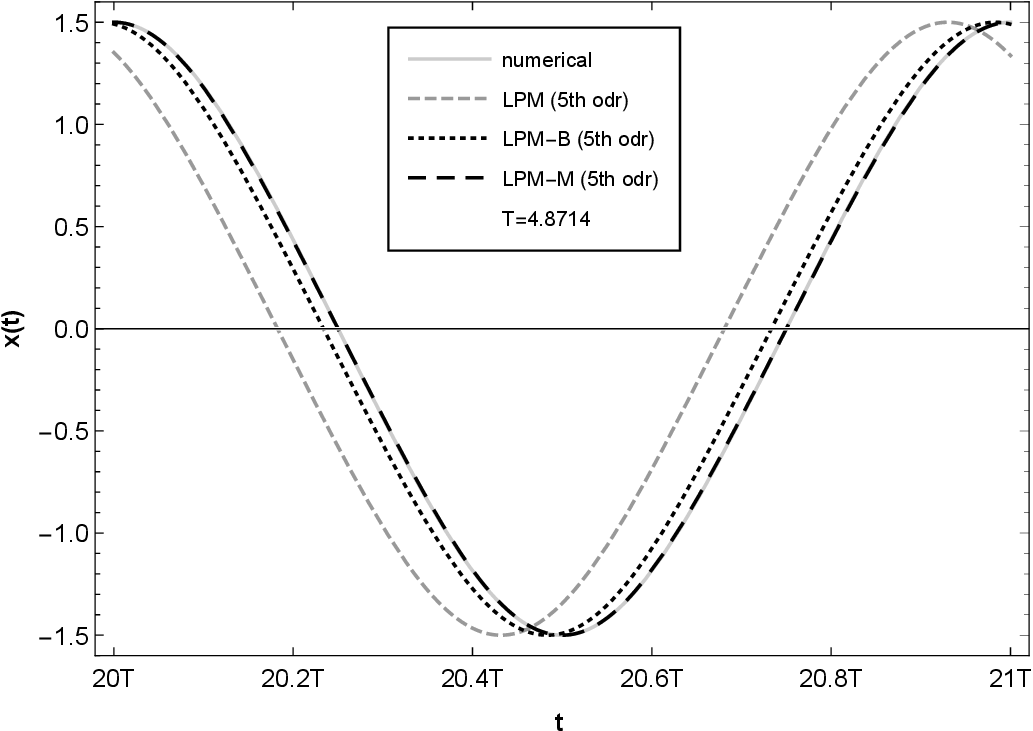}
\captionof{figure}{Comparison of $x_{[5]}(t)$ at higher time $(20T\leq t \leq 21T)$}
\label{fig:4}
\end{figure}

Moreover, the standard LPM and LPM-B solutions lag in phase for odd orders(Figures \ref{fig:2} and \ref{fig:3}) while for even orders the solutions lead in phase as is shown in Figure \ref{fig:4}.

\begin{figure}[h]
	\centering
    \includegraphics[width=.9\linewidth]{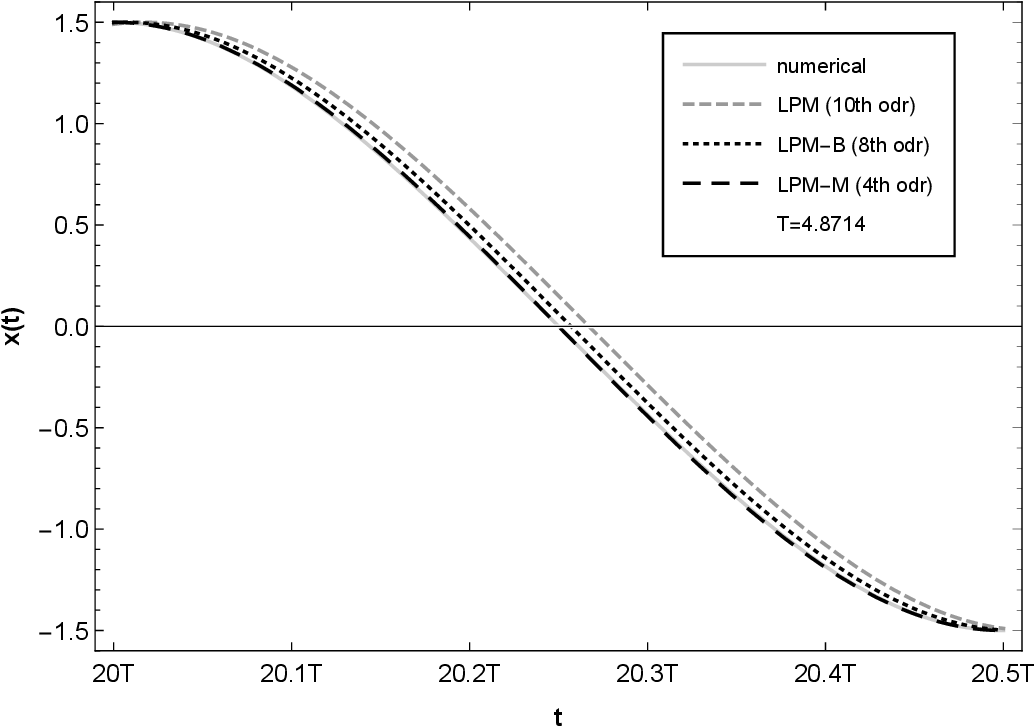}
\captionof{figure}{$x(t)$ in 10th order with LPM, 8th order with LPM-B and 4th order with LPM-M for $(20T\leq t \leq 20.5T)$}
\label{fig:5}
\end{figure}

In Figure \ref{fig:5}, the 4th order solution $x(t)$ obtained with LPM-M is shown together with the 8th order and 10th order solutions with LPM-B and LPM respectively. Convergence of LPM-M at 4th order itself, is significantly better than 8th order LPM-B which in turn is better than 10th order LPM.
Comparison of the velocities for the three methods shows a similar trend, confirming the proposed LPM-M's superior convergence.

\begin{figure}[h]
	\centering
    \includegraphics[width=.9\linewidth]{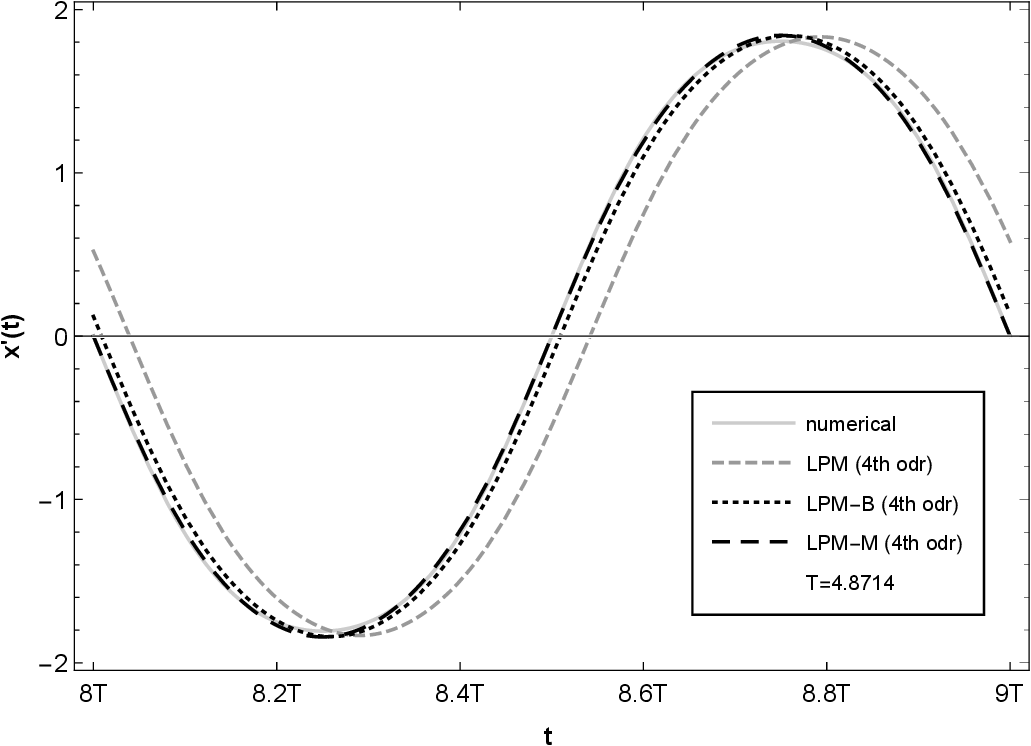}
\captionof{figure}{Comparison of $x_{[4]}'(t)$ with the three methods for time  $(8T\leq t \leq 9T)$}
\label{fig:6}
\end{figure}

\section{Conclusions:}
In this work, a modified Lindstedt–Poincaré method has been proposed to analyze the undamped Duffing oscillator. This method provides significant improvement over  the conventional Lindstedt–Poincaré method and Burton's improved Lindstedt–Poincaré method. The solutions obtained by LPM-M, shows excellent convergence with high precision numerical solution by 3rd order itself. For LPM-B and LPM similar agreement requires calculations upto at least 7th and 9th orders respectively. In addition, with LPM-M, there is no phase lag or gain over both small time and large time regions. Solutions by LPM-B and LPM both show phase shifts which become more pronounced as time increases. The proposed LPM-M method is simple to apply and is highly versatile, offering an effective analytical tool for studying a wide variety of nonlinear oscillators.

\section{Acknowledgements:}
RA acknowledges University Grants Commission, India for support through Junior Research Fellowship in Sciences (ID 521250).

\bibliographystyle{unsrt}   
\bibliography{references}

@book{strogatz2001nonlinear,
  title={Nonlinear dynamics and chaos: with applications to physics, biology, chemistry, and engineering (studies in nonlinearity)},
  author={Strogatz, Steven H},
  volume={1},
  year={2001},
  publisher={Westview press}
}

@book{nayfeh2024perturbation,
  title={Perturbation methods},
  author={Nayfeh, Ali H},
  year={2024},
  publisher={John Wiley \& Sons}
}

@book{nayfeh2024nonlinear,
  title={Nonlinear oscillations},
  author={Nayfeh, Ali H and Mook, Dean T},
  year={2024},
  publisher={John Wiley \& Sons}
}

@article{cveticanin2018strong,
  title={Strong nonlinear oscillators},
  author={Cveticanin, Livija},
  journal={Mathematical Engineering. Analytical Solution; Springer: Cham, Switzerland; Berlin, Germany},
  pages={1--296},
  year={2018},
  publisher={Springer}
}

@book{duffing1918erzwungene,
  title={Erzwungene Schwingungen bei ver{\"a}nderlicher Eigenfrequenz und ihre technische Bedeutung},
  author={Duffing, Georg},
  number={41-42},
  year={1918},
  publisher={Vieweg}
}

@article{feng2009homotopy,
  title={Homotopy analysis approach to Duffing-harmonic oscillator},
  author={Feng, Shao-dong and Chen, Li-qun},
  journal={Applied Mathematics and Mechanics},
  volume={30},
  number={9},
  pages={1083--1089},
  year={2009},
  publisher={Springer}
}

@article{momeni2011application,
  title={Application of He's energy balance method to Duffing-harmonic oscillators},
  author={Momeni, M and Jamshidi, N and Barari, Amin and Ganji, Davood Domiri},
  journal={International Journal of Computer Mathematics},
  volume={88},
  number={1},
  pages={135--144},
  year={2011},
  publisher={Taylor \& Francis}
}

@article{xu2007he,
  title={He's parameter-expanding methods for strongly nonlinear oscillators},
  author={Xu, Lan},
  journal={Journal of Computational and Applied Mathematics},
  volume={207},
  number={1},
  pages={148--154},
  year={2007},
  publisher={Elsevier}
}

@book{jordan2007nonlinear,
  title={Nonlinear ordinary differential equations: an introduction for scientists and engineers},
  author={Jordan, Dominic and Smith, Peter},
  year={2007},
  publisher={Oxford University Press}
}

@article{burton1984perturbation,
  title={A perturbation method for certain non-linear oscillators},
  author={Burton, TD},
  journal={International Journal of Non-Linear Mechanics},
  volume={19},
  number={5},
  pages={397--407},
  year={1984},
  publisher={Elsevier}
}

@article{Lindstedt1883,
  author  = {A. Lindstedt},
  journal = {Mem. Acad. Imper. Sci. St. Petersburg},
  volume  = {31},
  year    = {1883}
}

@book{abramowitz1964handbook,
  title={Handbook of Mathematical Functions with Formulas, Graphs, and Mathematical Tables},
  author={Abramowitz, Milton and Stegun, Irene A.},
  volume={55},
  year={1964},
  publisher={Dover Publications},
  address={New York},
  note={Formula 17.4.17}
}

\end{document}